\documentclass[12pt]{article}
\usepackage{epsf}

\newcommand{\VEV}[1]{\left\langle{#1}\right\rangle}
\renewcommand{\bar}[1]{\overline{#1}}

\begin {document}
\begin{flushright}
{\small
SLAC--PUB--8809\\
April 2001\\}
\end{flushright}


\normalsize

\vfill

\begin{center}
{{\bf\LARGE
QCD Aspects of Exclusive B Decays}\footnote{Work supported by
Department of Energy contract  DE--AC03--76SF00515.}}

\bigskip
S. J. Brodsky\\
{\sl Stanford Linear Accelerator Center, Stanford, CA 94309, USA\\
E-mail: sjbth@slac.stanford.edu}
\end{center}

\vfill

\begin{center}
{\bf\large
Abstract }
\end{center}

Exclusive $B$ decays can be factorized as convolutions of hard
scattering amplitudes involving the weak interaction with universal hadron
distribution amplitudes, thus providing a new QCD-based phenomenology.  In
addition,  semi-leptonic decay amplitudes can be computed exactly in
terms of the diagonal and off-diagonal
$\Delta n = 2$ overlap of hadronic light-cone wavefunctions.  I review
these formalisms and the essential QCD ingredients.  A canonical form of
the light-cone wavefunctions, valid at low values of the transverse
momenta, is presented.  The existence of intrinsic charm Fock states in
the $B$ meson wavefunction can enhance the production of final states of
$B$-decay with three charmed quarks,  such as $B \to J/\psi D \pi,$
as well as lead to the breakdown of the CKM hierarchy.

\vfill

\begin{center}
{ Invited talk presented at  \\4th International Conference on B Physics
and CP Violation (BCP4)} \\
{ Ago Town, Mie Prefecture, Japan }\\
{ 19--23 February 2001}\\
\end{center}

\vfill

\newpage

\section{Introduction}
Remarkable progress has recently been made applying
perturbative QCD methods to the exclusive
two-body hadronic decays of $B$ mesons.  Two groups,  Beneke,
Buchalla, Neubert, and Sachrajda 
\cite{Beneke:2000ry}, and Keum, Li, and Sanda \cite{Keum:2000wi}
have proven factorization theorems which allow the
rigorous computation of certain types of exclusive $B$ decay amplitudes in
terms of the distribution amplitudes of the initial-state $B$ meson and
the final-state hadrons.  These new analyses allow one to understand the
hadronic physics of heavy hadron decays from a fundamental
perspective.  There have been many applications of the PQCD formalism,
including many new results presented at this conference.

\begin{figure}[htb]
\center
{\epsfxsize=5.5in\epsfbox{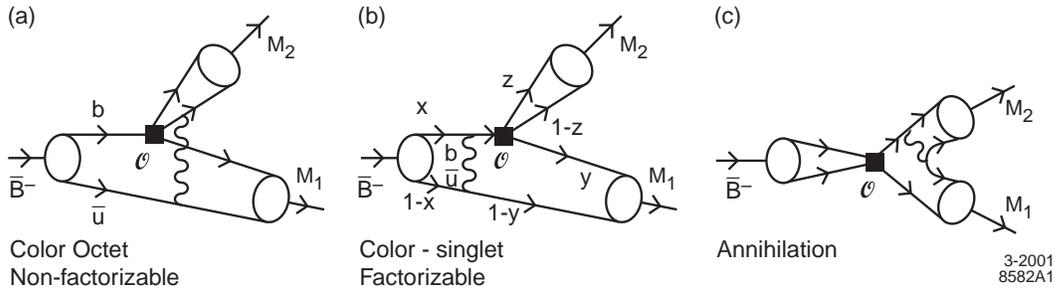}
\caption{Three representative contributions to exclusive $B$ decays to
meson pairs in PQCD.  The operators $\cal O$ represent the QCD-improved
effective weak interaction}
\label{fig:1}}
\end{figure}

As an example, consider the three representative contributions to the
decay of a $B$ meson to meson pairs illustrated in Fig. \ref{fig:1}.  In
Fig. \ref{fig:1}(a) the weak interaction effective operator ${\cal O}$
produces a $ q \bar q$ in a color octet state.  A gluon with virtuality 
$Q^2 = {\cal O} (M_B^2)$ is needed to equilibrate the large
momentum fraction carried by the $b$ quark in the $\bar B$ wavefunction.
The amplitude then factors into a hard QCD/electroweak subprocess
amplitude for quarks which are collinear with their respective
hadrons:
$T_H([b(x)
\bar u(1-x)]
\to [q(y) \bar u(1-y)]_1 [q(z) \bar q(1-z)]_2)$
convoluted with the distribution amplitudes $\phi(x,Q)$ \cite{Lepage:1980fj} of the
incident and final hadrons:
$${\cal M}_{octet}(B \to M_1 M_2) = \int^1_0 dz \int^1_0 dy \int^1_0 dx$$
$$\phi_B(x,Q) T_H(x,y,z) \phi_{M_1}(y,Q) \phi_{M_2}(z,Q).$$
Here $x = k^{+}/p^{+}_H = (k^0+ k^z) /(p^0_H + p^z_H)$ are the
light-cone momentum fractions carried by the valence quarks.

There are a several features of QCD which are required to
ensure the consistency of the PQCD approach: (a) the 
effective QCD coupling
$\alpha_s(Q^2)$ needs to be under control at the relevant scales of $B$ decay; 
(b) the distribution amplitudes of the hadrons need 
to satisfy convergence properties at the endpoints; and (c) one requires the 
coherent cancelation of the couplings of soft gluons to color-singlet states. 
This property, color transparency \cite{Brodsky:1988xz}, is a fundamental
coherence property of gauge theory and leads to diminished final-state
interactions and corrections to the PQCD factorizable contributions.  Color transparency
is not in contradiction with Watson's theorem for the factorization
of final state phases since the energy in $B$ decay is well above the
regime where the scattering of the final states is
elastic \cite{Suzuki:1999uc}.

An innovative experiment (E791) at Fermilab has recently reported a
direct determination of the quark-antiquark pion light-cone wavefunction
by measuring the momentum distribution of diffractive dijet production
on nuclei, $\pi A \to {\rm jet ~ jet}~ A$, where the final nucleus stays
intact \cite{Aitala:2000hb}.  The measured pion wavefunction appears to be similar
to the form $\phi_M(x) \propto x(1-x)$ which is the asymptotic solution
to the evolution equation for the pion distribution amplitude.  

The E791 experiment \cite{Aitala:2000hc} has also provided
a remarkable confirmation of color transparency. Color transparency implies that when
a fast hadron fluctuates into configurations of small transverse size, it can
interact weakly and thus transverse a nucleus with minimal
interactions \cite{Bertsch:1981py,Frankfurt:2000jm}.
E791 finds that the diffractive dijet
production process occurs coherently on each nucleon in the
nucleus without nuclear absorption when the jet transverse momentum
exceeds $2$ GeV/c.  The EVA spectrometer experiment E850
\cite{Leksanov:2000hm} at Brookhaven has also reported striking effects
of color transparency in quasi-elastic proton-proton scattering on nuclei.

\begin{figure}[htb]
\center
{\epsfxsize=5.5in\epsfbox{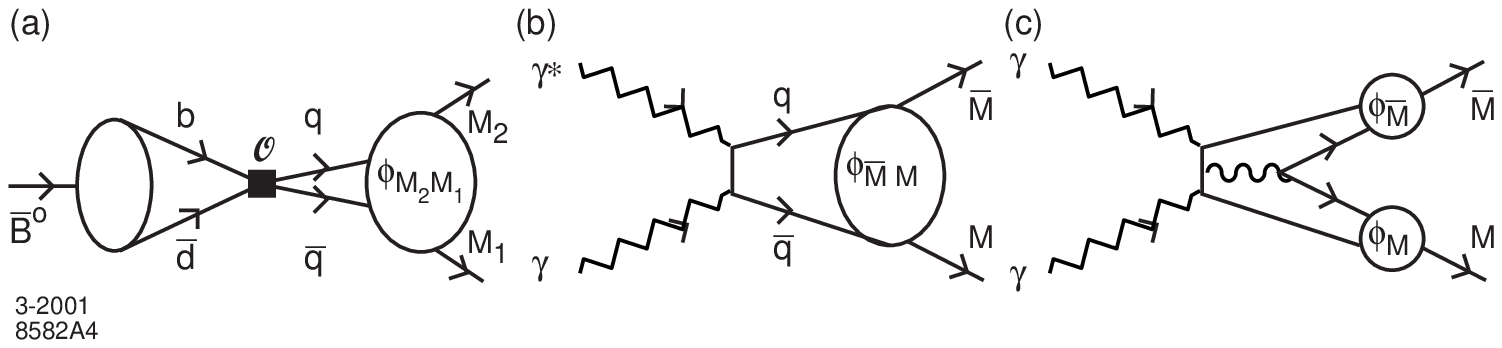}
\caption{Similarities between exclusive $B \to M_1 M_2$ decays and $\gamma
\gamma \to M \bar M$ decays in PQCD.  }
\label{fig:aa}}
\end{figure}

The distribution amplitudes
$\phi_H(x,Q)$, which control hard exclusive $B$ decays are fundamental
gauge-invariant hadronic wavefunctions  \cite{Lepage:1980fj}. These
amplitudes describe the light-cone momentum distributions of the valence
quarks with relative transverse momentum
$k_\perp < Q$.  Evolution equations can be derived within PQCD which
describe the change of $\phi(x,Q)$ with respect to $\ell n Q^2$.  They can
also be expanded in terms of specific polynomial eigenfunctions determined
by conformal symmetry and corresponding to local operators with given
anomalous dimensions   \cite{Brodsky:1980ny,Braun:1990iv}.  The distribution
amplitude can also be understood as the transverse momentum integral of
the lowest Fock state of the light-cone wavefunction of the hadron
$\psi(x_i,k_{\perp i}),$ or equivalently it can be computed from the
Bethe-Salpeter amplitude evaluated at equal light-cone time
$\tau = t + z/c$ in light-cone gauge $A^+ = 0.$ Typically, one is
interested in the leading twist-2 amplitude where the helicities of
the pseudoscalar or vector mesons have opposite light-cone helicity.
Since they are process independent, the distribution amplitudes can be
measured in other high momentum exclusive reactions, such as
$\gamma \gamma \to M \bar M.$   See Fig. \ref{fig:aa}.  One can also 
make a connection between
exclusive $B \to M_1 M_2$ amplitudes with the di-meson distribution
amplitudes of meson pairs determined by measurements of $\gamma^* \gamma
\to \pi^+ \pi^-$ \cite{Diehl:2000uv}.

The
endpoint-regions of integration $x \to 1$, $y \to 1$, $z \to 1$
of the octet
amplitude converge because of the sufficiently fast fall off of
the distribution amplitudes.  This follows from the form of the
solutions of the QCD evolution equation and also from the requirement
that the light-cone kinetic energy of a quark
$(k^2_\perp + m^2_\perp )/ k^+$ has finite expectation value.  Note also
that $x \to 1$ or
$ x \to 0$ corresponds kinematically to $k_z \to - \infty$ in the rest
frame wavefunction. The contributions of 
higher Fock states containing extra gluons or sea quarks 
are suppressed by powers of
$m_b^{-1}$ in light-cone gauge.  Soft gluon corrections such as those
leading to final-state interactions are suppressed by the color
neutrality of the final state hadrons as they emerge with small color
dipole moments.  Insertions of collinear gluons from higher Fock states
into the hard amplitude are power-law suppressed in light-cone gauge 
\cite{Srivastava:2000cf}.  The proofs of factorization for exclusive $B$
decays are generalizations of the analyses of exclusive amplitudes involving
large momentum transfer 
\cite{Lepage:1980fj,Chernyak:1984ej}.

Can one trust the applicability of leading twist PQCD
factorization theorems at momentum transfers involved in $B$ decays?  The CLEO collaboration
\cite{Gronberg:1998fj} has verified the scaling and angular predictions
for hard exclusive two-photon processes such as $\gamma^* \gamma \to
\pi^0$ and $\gamma \gamma \to \pi^+ \pi^-$ at momentum transfers
comparable to the momenta that occur in exclusive $B$-decays.
The CLEO data
for the sum of $\gamma
\gamma \to \pi^+ \pi^+ + K^+ K^-$ channels at $W = \sqrt s > 2. 5$ GeV are
in striking agreement with the perturbative QCD predictions.  Moreover,
the observed angular distribution shows a striking transition to the predicted QCD
form as $W$ is raised.  The L3
experiment at LEP at CERN
\cite{Vogt:1999qd} has also measured a number of exclusive hadron
production channels in two-photon processes, providing important
constraints on baryon and meson distribution amplitudes and checks of
perturbative QCD factorization.
The $\gamma^* \gamma \to \pi^0$ results
are in close agreement with the scaling and normalization of the PQCD
prediction, provided that the pion distribution amplitude
$\phi_\pi(x,Q)$ is close to the $x(1-x)$ form, the asymptotic solution
to the evolution equation, in agreement with dijet diffraction
measurements from E791.  It will be important to have measurements of
processes such as $\gamma \gamma \to \pi^0 \pi^0$ and $\gamma \gamma
\to \rho^+ \rho^-$ since they are particularly sensitive to the shape of
meson distribution amplitudes  \cite{Brodsky:2000dq}.

The scale of the coupling which controls
$T_H$ in hard exclusive processes can be obtained using the BLM method;
the scale is of order of the gluon virtuality and the perturbative coefficients are 
identical to those of conformal
theory \cite{Brodsky:1983gc,Brodsky:2000cr}.  The empirical success of
dimensional counting rules for exclusive two-photon reactions and other
exclusive amplitudes at moderate momentum transfer suggests that
the QCD coupling $\alpha_s(Q^2)$ is effectively constant at low $Q^2$ 
\cite{Brodsky:1998dh}.The effective
charge
$\alpha_{\tau}(s)$ measured in hadronic $\tau$ decays integrated up to
invariant mass squared $s$ has been shown to have this property 
\cite{Sven}.   All of these aspects of QCD are essential for the analysis of
exclusive
$B$ decays.

In contrast to the color-octet contribution, the contribution of
Fig. \ref{fig:1}(b) to the $B
\to M_1 M_2$ amplitude where the light quarks of meson $M_2$ are produced
in a color-singlet state has an apparent $(1-y)^{-2}$ endpoint
singularity in the limit of large heavy quark mass 
\cite{Szczepaniak:1990dt}.  Thus one cannot guarantee the convergence of
the convolution of $T_H$ solely from the fall-off of the
distribution amplitudes; thus strictly speaking such amplitudes cannot be
analyzed at fixed order in PQCD.  The simplest approach, taken by
Buchalla {\it et al.}, is to identify the required color-singlet
transition amplitude $B \to M_1$ with the transition form factor $F_{B
\to M_1 \ell \bar \nu}(M^2_2)$ measurable in semileptonic decays.

Despite the complication from the endpoint sensitivity of the color-singlet contributions, some
predictability is still possible from PQCD.  Li {\it et al.}
\cite{Li:2001ay,Keum:2001ms} have noted that
Sudakov form factors control will effectively suppress any power-law
endpoint singularities which appear when the quarks have finite transverse
momentum.  This type of Sudakov suppression is also an essential element in the
analysis of the $x \to 1$ endpoint, finite $k_T,$ contributions to the
elastic form factors of nucleons \cite{Lepage:1980fj}. Physically, such
Sudakov effects occur when a quark near its mass shell is forced to
decelerate without gluonic radiation.  The resummation of a
double-logarithmic series gives the form ($C_F = {N_C^2 - 1 \over 2
N_C}$):
$S(Q^2) = \exp{-{\alpha_s C_F \over 4 \pi}\ell n^2 {Q^2\over k^2_\perp}}$ at
fixed QCD coupling
which suppresses the long-distance
contributions in the large $b$ region. The exponent becomes a
$\ell n \times \ell n \ell n$ form when the running of the QCD coupling is taken into account.  
Keum and Li  \cite{Keum:2001ms}
find that almost all of the contribution to the exclusive matrix elements then 
come from the integration region where $\alpha_s/\pi < 0.3 $ and the  perturbative 
QCD treatment can be is justified. Thus for PQCD contributions which are 
Sudakov form factor controlled, the effective scale of the subprocess is
$Q^2 = {\cal O} (m_b \Lambda_{QCD}).$  Li {\it et al.} argue that the presence of
Sudakov suppression allows the applicability of PQCD to exclusive
decays.

The Sudakov suppression approach also allows new predictions for the
phase structure of the annihilation graphs such as Fig. \ref{fig:1}(c).
In this case, the endpoint behavior of the higher twist wavefunctions
which appear in the helicity parallel amplitudes as determined by QCD
evolution may not be sufficient to control the endpoint singularity
region.  Such amplitudes, however, would be controlled by the Sudakov form
factors.  Since the scale of the exchanged gluon momentum is of order 
$\Lambda_{QCD} m_b$, the coupling strength of the exchanged gluons would
be enhanced by a larger effective
$\alpha_s(Q^2),$ thus leading to the possibility of enhanced
$CP$-violating phases  \cite{Keum:2001ms}. 
Further discussion of the phenomenology of exclusive QCD decays based
on the factorization theorems and the Sudakov-extended analyses has been 
given by Li  \cite{Li:2001ay}, Neubert  \cite{Neubert:2000ag}, and Becher 
{\it et al.} \cite{Becher:2001hu}  It should also be
possible to extend the PQCD formalism to the exclusive decays of
$B$ mesons to baryon pairs.

\section{Exact Representation of Electroweak Matrix Elements in terms of
Light-Cone Wavefunctions}

\begin{figure}[htb]
\center
{\epsfxsize=5.2in\epsfbox{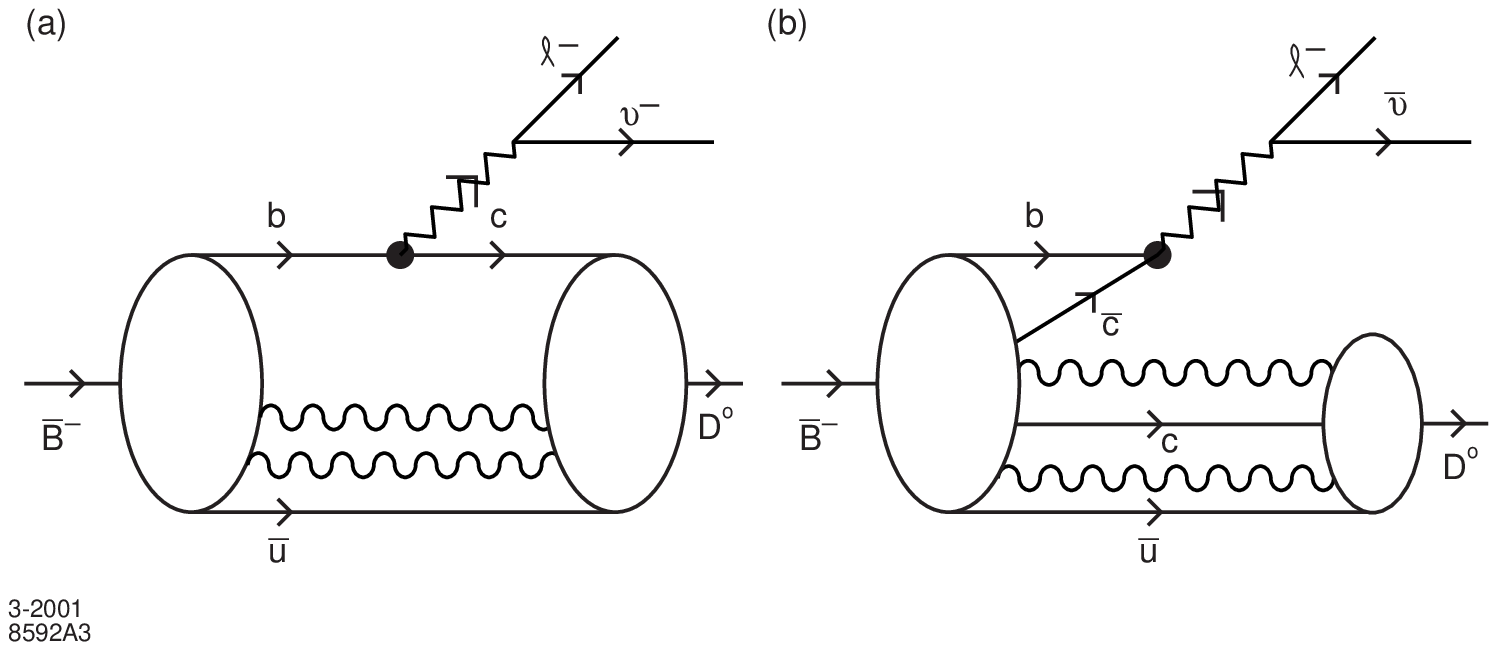}
\caption{Diagonal and off-diagonal contributions to time-like $B$ decays.
Both contributions are required by Lorentz covariance.}
\label{fig:bb}}
\end{figure}

The natural calculus for describing the
bound-state structure of relativistic composite systems in quantum field
theory is the light-front Fock expansion which encodes the properties of
a hadrons in terms of a set of frame-independent $n$-particle
wavefunctions.
A remarkable advantage of the light-cone formalism is that
exclusive semileptonic
$B$-decay amplitudes such as $B\rightarrow A \ell \bar{\nu}$ can be
evaluated {\it exactly} \cite{Brodsky:1999hn}.
The time-like decay matrix elements require the computation of the
diagonal matrix element $n \rightarrow n$ [See Fig. \ref{fig:bb}(a)] where
parton number is conserved, and the off-diagonal $n+1\rightarrow n-1$
convolution where the current operator annihilates a $q{\bar{q'}}$ pair
in the initial $B$ wavefunction, as illustrated in Fig. \ref{fig:bb}(b).
This term is a consequence of the
fact that the time-like decay $q^2 = (p_\ell + p_{\bar{\nu}} )^2 > 0$
requires a positive light-cone momentum fraction $q^+ > 0$.  The
sum of amplitudes is required to obtain a covariant result.  A similar
form also controls deeply virtual Compton scattering.  In contrast,
for space-like currents, one can choose $q^+=0$, as in the
Drell-Yan-West representation of the space-like electromagnetic form
factors.

The light-cone Fock expansion is defined as the projection of an exact
eigensolution of the full light-cone quantized Hamiltonian on the
solutions of the free Hamiltonian with the same global quantum numbers.
The
coefficients of the Fock expansion are the complete set of $n$-particle
light-cone wavefunctions, $\{\psi_n(x_i, k_{\perp i}, \lambda_i)\}$.
The coordinates $x_i, k_{\perp i}$ are internal relative coordinates,
independent of the total momentum of the bound state, and satisfy $ 0 <
x_i < 1,\ \sum_i^n x_i = 1$ and $\sum_i^n k_{\perp i} = 0_\perp$.  The light-cone 
wavefunctions are thus independent of the total hadronic momentum.

\begin{figure}[htb]
\center
\leavevmode
{\epsfysize=4in\epsfbox{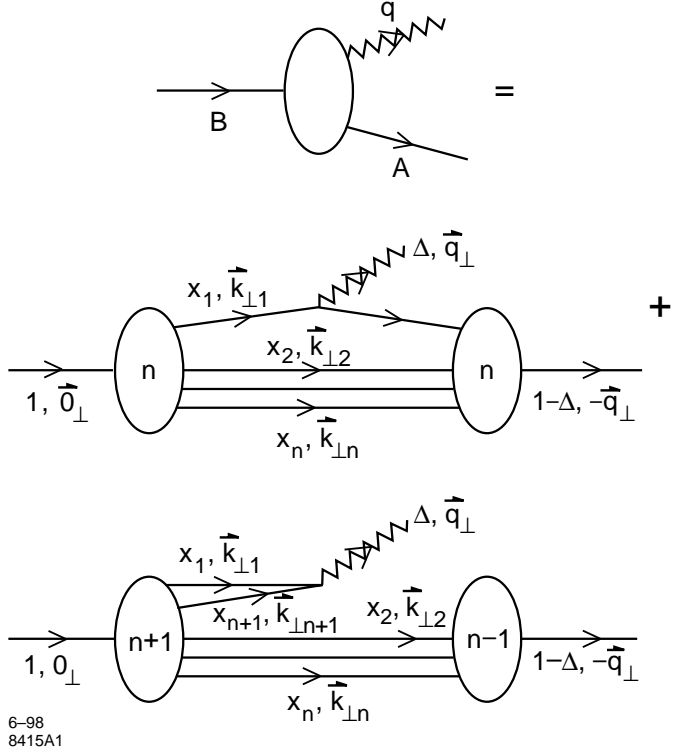}
\caption[*]{Exact representation of electroweak decays
and time-like form factors in the light-cone Fock representation.}
\label{fig:cc}}
\end{figure}

The evaluation of the semileptonic decay amplitude $B \to A \ell {\bar{\nu}}$
requires the matrix element of the weak current between hadron states
$\VEV{A \vert j^\mu(0) \vert B}$.  (See Fig.~\ref{fig:cc}.)
We can choose the Lorentz frame where the outgoing
leptonic current carries
$q^\mu = \left(q^+, q_\perp, q^- \right)= \left(\Delta P^+, q_\perp,
{q^2+q^2_\perp\over \Delta P^+}\right)$.
In the limit $\Delta \to 0$, the
matrix element for the $+$ vector current coincides with the Drell-Yan
formula for spacelike form factors.

\vspace{.5cm}

The diagonal (parton-number-conserving) matrix
element has the form of a convolution of $\psi^\dagger_{A (n)}(x^\prime_i, 
{\vec{k}}^\prime_{\perp i},\lambda_i)$ and 
$\psi_{B (n)}(x_i, {\vec{k}}_{\perp i},\lambda_i)$
where
$x^\prime_1 = {x_1-\Delta \over 1-\Delta}\, ,\
{\vec{k}}^\prime_{\perp 1} ={\vec{k}}_{\perp 1}
- {1-x_1\over 1-\Delta} {\vec{q}}_\perp$
for the struck quark and 
$x^\prime_i = {x_i\over 1-\Delta}\, ,\
{\vec{k}}^\prime_{\perp i} ={\vec{k}}_{\perp i}
+ {x_i\over 1-\Delta} {\vec{q}}_\perp$ for the $ (n-1)$ spectators.
If quark masses are neglected the vector and
axial currents conserve helicity.
Note that $\sum_i^n x^\prime_i = 1$, and
$\sum_i^n {\vec{k}}^\prime_{\perp i} = {\vec{0}}_\perp$.

For the $n+1 \to n-1$ off-diagonal term,
consider the case where
partons $1$ and
$n+1$ of the initial wavefunction annihilate
into the leptonic current leaving
$n-1$ spectators.
Then $x_{n+1} = \Delta - x_{1}$,
${\vec{k}}_{\perp n+1} = {\vec{q}}_\perp-{\vec{k}}_{\perp 1}$.
The remaining $n-1$ partons have total momentum
$((1-\Delta)P^+, -{\vec{q}}_{\perp})$.
The final wavefunction then has arguments
$x^\prime_i = {x_i \over (1- \Delta)}$ and
${\vec{k}}^\prime_{\perp i}=
{\vec{k}}_{\perp i} + {x_i\over 1-\Delta} {\vec{q}}_\perp .$
The arguments of the final-state wavefunction
satisfy
$\sum_{i=2}^n x^\prime_i = 1$,
$\sum_{i=2}^n {\vec{k}}^\prime_{\perp i} = {\vec{0}}_\perp$.

The off-diagonal $n+1 \rightarrow n-1$ contributions give a new
perspective for the physics of $B$-decays.  A semileptonic decay
involves not only matrix elements where a quark changes flavor, but also
a contribution where the leptonic pair is created from the annihilation
of a $q {\bar{q'}}$ pair within the Fock states of the initial $B$
wavefunction.  The semileptonic decay thus can occur from the
annihilation of a nonvalence quark-antiquark pair in the initial hadron.
This feature will carry over to exclusive hadronic $B$-decays, such as
$B^0 \rightarrow \pi^-D^+$.  In this case the pion can be produced from
the coalescence of a $d\bar u$ pair emerging from the initial higher
particle number Fock wavefunction of the $B$.  The $D$ meson is then
formed from the remaining quarks after the internal exchange of a $W$
boson.

In principal, the evaluation of the hadronic matrix elements
needed for $B$-decays and other exclusive electroweak decay amplitudes
requires knowledge of all of the light-cone Fock wavefunctions of the
initial and final state hadrons. 
In practice only the first few Fock components have significant magnitude. 
In the case of model gauge theories
such as QCD(1+1)$\,$\cite{Horn} or collinear QCD$\,$\cite{AD} in one-space
and one-time dimensions, the complete evaluation of the light-cone
wavefunction is possible for each baryon or meson bound-state using the
DLCQ method.  In fact, solutions of the t'Hooft model have been used to
model physical $B$-decays 
\cite{Grinstein:1999gc,Lebed:2000gm,Bigi:1999kc,Neubert:2000ux}.

\section {Canonical Form for Light-Cone Wavefunctions at Small Transverse
Momenta}

One can understand some basic features of the LC wavefunctions of hadrons, including the
wavefunction of the $B$ meson itself by considering its canonical form at
low transverse moments $k_{\perp i}$: \cite{DSH}
$$\psi_n(x_i,k_{\perp i}) = {\Gamma(x_i,k_{\perp i}) \over [\delta^2 +
\sum^n_{j=1} {(x_j - \hat x_j)^2\over  x_j}]}$$
where $\delta^2 = 2B.E./M_H$,
$\hat x_j = {m_{\perp j}/ \sum^n_{k=1} m_{\perp k}}$, 
and the transverse mass squared  is $m^2_{
\perp j} = m_j^2 + k^2_{ \perp j}.$ Here $\Gamma$
is the convolution of the interaction kernel $V_{LC}$ with the light-cone
wavefunction and can be considered as slowly varying compared to the
peaking of the denominator.  The spin structure of the numerator $\Gamma$ 
can be largely constructed using light-cone angular momentum conservation 
\cite{Brodsky:2001ii}.
The canonical form follows
directly from the bound-state light-cone Hamiltonian equation of motion
$\left[M^2 -
\sum^n_{i = 1} m^2_\perp\right] \psi = V_{LC} \Psi$.
One sees that at small values of
$k_{\perp i}$, the light-cone wavefunctions peak at value
$x_i = \hat x_i$, the equal rapidity, minimal off-shell configuration.
The spread of the wavefunction where the wavefunction falls to half of its 
value is 
$<(x_i - \hat x_i)^2> =  x_i
\delta^2$ and is largest for the
heaviest partons.   The full wavefunction can then be constructed from this low 
$k_\perp$ starting point by iterating the
equation of motion  \cite{Lepage:1980fj}.

DLCQ (discretized light-cone quantization) is a method which solves
quantum field theory by direct diagonalization of the light-front
Hamiltonian  \cite{Pauli:1985ps,Brodsky:1998de}.  The DLCQ method has had
much success in solving quantum field theories in low space-time
dimensions and has also found utility in string theory.  There has been
progress recently in systematically developing the computation and
renormalization methods needed to make DLCQ viable for QCD in physical
spacetime.  There has also been considerable success in applying these
methods to simple quantum field theories in 3+1 dimensions 
\cite{Brodsky:1999xj}.  The distribution amplitude of the pion has recently
been computed using a combination of the discretized light-cone
quantization and transverse lattice methods.
\cite{Dalley:2000dh}

\section{Intrinsic Charm and B Decay}

The complete Fock state description of the $B$ meson wavefunction
contains higher particle number states such as $\vert b \bar u g>$,
$\vert b \bar u s \bar s>$, and $\vert b \bar u c \bar c>.$ Such Fock
components arise from gluon splitting and from diagrams in which the sea
quarks are multi-connected to the valence quarks.  Since the light-cone
wavefunction peaks at $x_i \propto m_{i \perp},$ the intrinsic charm
quarks are found at relatively high $x$  \cite{Brodsky:1980pb}.  Evidence
for a $1\%$ probability of intrinsic charm in the proton has been given
by Harris {\it et al.} \cite{Harris:1996jx} based on an analysis of the
EMC measurement of the proton's charm structure function at large $x$.
Intrinsic charm can also explain the $J/\psi \to \rho \pi$ puzzle 
\cite{Brodsky:1997fj}.  Franz {\it et al.} \cite{Franz:2000ee} have
recently given a rigorous operator product expansion of intrinsic heavy
quarks showing that the momentum fraction carried by intrinsic heavy quarks 
$,x_{Q \bar Q}$  scales as $p_H^2/m_Q^2$ where $p_H$ is the
characteristic internal momentum in the hadron.  This is strictly a non-Abelian
effect and only occurs for intrinsic charm pairs in a color octet
configuration  \cite{Gardner}.  Since the $B$ wavefunction is more compact
compared to light hadrons (a reduced mass effect), the magnitude
intrinsic charm is dynamically enhanced in the $B$ wavefunction.

\begin{figure}[htb]
\center
{\epsfxsize=5.5in\epsfbox{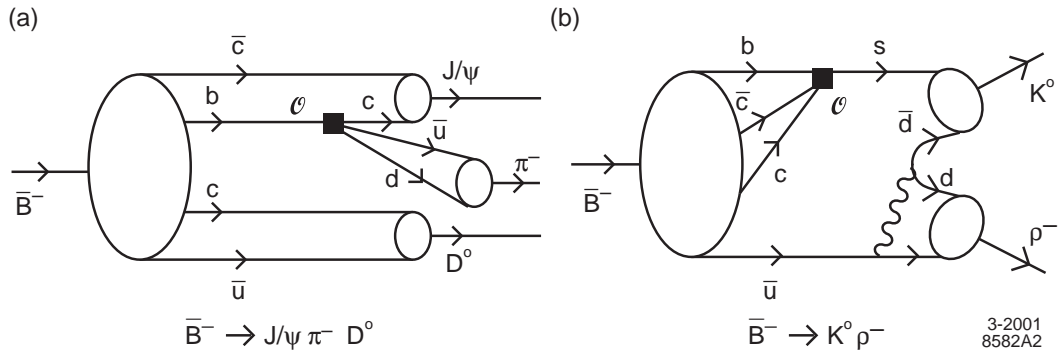}
\caption{Examples of intrinsic charm effects in exclusive $B$ decays.  In
(a) the intrinsic charm of the $B$ wavefunction leads to final states
with three charmed quarks.  In (b), the annihilation of the intrinsic
charm quarks in the effective weak interaction allows a decay to proceed
without CKM suppression.}
\label{fig:3}}
\end{figure}

The presence of intrinsic charm quarks in the $B$ wavefunction provides
new mechanisms for $B$ decays.  For example, Chang and
Hou \cite{Chang:2001iy} have considered the production of final states
with three charmed quarks such as $B \to J/\psi D \pi$ and $B \to J/\psi
D^*$ which arises naturally when the $b$ quark of the intrinsic charm
Fock state $\vert b \bar u c \bar c>$ decays $b \to c \bar u d$.  See
Fig. \ref{fig:3}(a).  In fact, the $J/\psi$ spectrum for inclusive $B \to J/\psi
X$ decays measured by CLEO and Belle does show a distinct enhancement at
low $J/\psi$ momentum where such decays would occur  \cite{Chang:2001iy}.
Another interesting possibility is that the enhancement at low $J/\psi$
momentum is due to final states containing baryon pairs.  $B
\to J/\psi
\bar p
\Lambda$ distribution which would also allow the study of di-baryons or
the bound states of baryons with the
$J/\psi$ which can arise via the QCD van der Waals
potential at low relative velocity  \cite{Brodsky:1997yr}.

The existence of intrinsic charm in the $B$ wavefunction, even at a few
percent probability, allows exclusive decays of the B meson which
evade the hierarchy of the CKM matrix \cite{Gardner}.  This can lead to
significant modifications of standard predictions for exclusive decays
such as $B
\to \rho K$.
In the standard analyses, the tree-level contributions
involving $b \to u \bar u s$ are CKM suppressed, whereas the
penguin contributions which are not CKM suppressed are 
numerically suppressed by Wilson coefficients. 

An example of an
intrinsic charm contribution to $B \to \rho \pi$ is illustrated in Fig.
\ref{fig:3}(b).  The three-particle weak annihilation process $b c \bar c
\to s g $ can occur via $W^-$ exchange tree-level operators using the
large
$V_{c b}$ and
$V_{cs}$ matrix element.  The non-valence quark
annihilation process thus provides a new mechanism for decays such as
$B^- \to \rho^0 K^-$.  The relatively small intrinsic charm probability of
$\sim 4 \%$ can thus be offset by the comparatively large CKM matrix
elements of the effective weak interaction.  Furthermore, since the same
initial and final states are involved, the intrinsic charm annihilation
amplitudes can interfere with the conventional amplitudes, leading to a
significant correction to the traditional analyses for decay amplitudes
involving the intrinsic charm Fock state.

\section{Conclusions}

Light-cone wavefunctions provide a fundamental frame-independent
description of hadrons in terms of their quark and gluon degrees of
freedom at the amplitude level and  a natural description of 
the exclusive  amplitudes which occur in $B$ decay.  For example, the semi-leptonic 
decay amplitudes of hadrons can be computed exactly in
terms of the diagonal $\Delta n = 0 $ and off-diagonal
$\Delta n = 2$ overlap of the hadronic light-cone wavefunctions.  Since the momentum 
transfers are large, exclusive $B$ decays into hadron pairs can be factorized as 
convolutions of hard
scattering amplitudes involving the weak interaction together with universal hadron
distribution amplitudes, thus providing a new perturbative QCD-based phenomenology.
These factorization theorems, together with the universality
of the distribution amplitudes, provide
a profound connection between hard scattering exclusive processes
such as elastic form factors, two-photon reactions, and heavy hadron
decays. We can understand the shape of the light-cone wavefunctions using
a canonical form which is  valid at low values of the transverse
momenta.
Color transparency 
implies minimal phases from final-state interactions, a fact which is critical to the
interpretation of $CP$ violation in $B$ physics.  
I have also emphasized the fact the existence of intrinsic charm Fock states in
the $B$-meson wavefunction can enhance the production of final states of
the $B$ with three charmed quarks,  such as $B \to J/\psi D \pi,$
as well as lead to the breakdown of the CKM hierarchy.

\section{Acknowledgments} I wish to thank Tony Sanda and his
colleagues at BCP4 for organizing this outstanding meeting.  I
also wish to thank  Alex Kagan, Yong Yeon Keum, Hsiang-nan  Li,
Sven Menke, Carlos Merino, Mathias Neubert, Helen Quinn,  and Lincoln 
Wolfenstein for helpful discussions. Parts of this work are based on 
collaborations with  Susan Gardner, and Dae Sung Hwang.

\end{document}